\documentclass[12pt]{article}
\usepackage[centertags]{amsmath}
\usepackage{amssymb,amsfonts}
\usepackage[comma,super]{natbib}
\usepackage{latexsym}
\begin{document}
\begin{center}
{\Large \bf Mixed potentials in radiative stellar collapse}\\
\vspace{1.5cm}
 {\bf S. Thirukkanesh\footnote{Permanent address:
Department of Mathematics, Eastern University, Chenkalady, Sri
Lanka.\\ Electronic mail: thirukkanesh@yahoo.co.uk} and S. D.
Maharaj\footnote{Electronic mail: maharaj@ukzn.ac.za}}\\
Astrophysics and Cosmology Research Unit,\\
School of Mathematical Sciences,\\
University of KwaZulu-Natal,\\
Private Bag X54001,\\
Durban 4000,\\
South Africa.\\

\vspace{1.5cm}

{\bf Abstract}\\
\end{center}
We study the behaviour of a radiating star when the interior
expanding, shearing fluid particles are traveling in geodesic
motion. We demonstrate that it is possible to obtain new classes of
exact solutions in terms of elementary functions without assuming a
separable form for the gravitational potentials or initially fixing
the temporal evolution of the model unlike earlier treatments. A
systematic approach enables us to write the junction condition as a
Riccati equation which under particular conditions may be
transformed into a separable equation. New classes of solutions are
generated which allow for mixed spatial and temporal dependence in
the metric functions. We regain particular models found previously
from our general classes of solutions.

\section{Introduction}
In an astrophysical environment, it is likely that a  star emits
radiation and particles in the process of gravitational collapse. In
this situation, the heat flow in the interior of a star should not
be neglected; the interior spacetime of the collapsing radiating
star should match to the exterior spacetime described by the Vaidya
\cite{1} solution. Exact models of relativistic radiating stars are
important for the investigation of the cosmic censorship hypothesis
and gravitational collapse \cite{2,3}. Santos \cite{4} formulated the junction
conditions for shear-free collapse, matching the interior metric
with the exterior Vaidya metric at the boundary of the star, which
made it possible to generate exact models. This treatment enabled us
to investigate physical features such as surface luminosity,
dynamical stability, relaxation effects, particle production at the
surface and temperature profiles for radiating stars in general
relativity. De Oliviera \emph{et al} \cite{5} proposed a radiating model
of an initial interior static configuration leading to slow
gravitational collapse. It had been shown earlier that the slowest
possible collapse arises in the case of shear-free fluid interiors
\cite{6}. In a recent treatment Herrera \emph{et al} \cite{7} proposed a
relativistic radiating model with a vanishing Weyl tensor, in a
first order approximation, without solving the junction condition
exactly. Then Maharaj and Govender \cite{8} and Herrera \emph{et al} \cite{9}
solved the relevant junction condition exactly, and generated
classes of solutions in terms of elementary functions which contain
the Friedmann dust solution as special case. Later Misthry \emph{et
al} \cite{10} obtained several other classes of solution by transforming
the junction condition to the form of an Abel equation of the first
kind. These exact models have proved to be useful in analysing the
relativistic behaviour of a collapsing objects in the stellar
scenario.

Another useful approach in studying the effects of dissipation is
due to Kolassis \emph{et al} \cite{11} in which the fluid particles are
restricted to travel along geodesics. In the absence of heat flow,
the interior Friedmann dust solution is regained. This particular
exact solution formed the basis for many investigations involving
physical features such as the rate of collapse, surface luminosity
and temperature profiles. The physical investigations include the
analytic model of radiating gravitational collapse in a spherical
geometry with neutrino flux by Grammenos and Kolassis \cite{12}, the
model describing realistic astrophysical processes with heat flux by
Tomimura and Nunes \cite{13}, and models undergoing gravitational
collapse with heat flow which serves as a possible mechanism for
gamma-ray bursts by Zhe \emph{et al} \cite{14}. Herrera \emph{et al} \cite{15}
investigated geodesic fluid spheres in coordinates which are not
comoving in the presence of anisotropic pressures. Govender \emph{et
al} \cite{16} demonstrated that the temperature in casual thermodynamics
for particles traveling on geodesics produces higher central values
than the Eckart theory. The first exact solution, that we are aware
of, with nonzero shear was obtained by Naidu \emph{et al} \cite{17},
considering geodesic motion of fluid particles; later Rajah and
Maharaj \cite{18} obtained two classes of nonsingular solutions by
solving a Riccati equation. Note that the geodesic condition arises
in several other astrophysical situations including Euclidean stars,
with the real and proper radius equal, in the absence of dissipation
\cite{19}.

In this paper we attempt to extend the previous treatments of Naidu
\emph{et al} \cite{17} and Rajah and Maharaj \cite{18} through a systematic
approach by studying the fundamental junction condition. Our
intention is to show that the nonlinear boundary condition may be
analysed mathematically  to produce an infinite classes of exact
solutions. In Section 2, we present the geodesic model governing the
description of a radiating star using the Einstein field equations
together with the junction conditions in the presence of anisotropic
pressure. We show that it is possible to transform the junction
condition into a separable equation by placing restrictions on one
of the gravitational potentials.  Two new classes of solutions are
obtained in terms of arbitrary functions of the radial coordinate
and we regain the models found in the past for particular choices of
arbitrary functions  in Section 3. In Section 4, we discuss some
physical aspects of the models generated.

\section{The model}
In the context of general relativity, the form for the interior
space time of a spherically symmetric collapsing star with
nonzero shear when the fluid trajectories are geodesics is given
by the line metric
\begin{equation}
\label{eq:g1} ds^2 =- dt^2 +B^2 dr^2 +Y^2(d\theta^2 +\sin^2 \theta
d\phi^2),
\end{equation}
where $B$ and $Y$ are functions of both the temporal coordinate
$t$ and radial coordinate $r$. The fluid four-velocity vector
$\mathbf{u}$ is given by $u^a ={\delta}_0^a$ which is comoving.
For the line element (\ref{eq:g1}), the four-acceleration
$\dot{u}^a$, the expansion scalar $\Theta$ and the magnitude of
the shear scalar $\sigma$ are given by
\begin{subequations}
\begin{eqnarray}
\label{eq:g2a}\dot{u}^a &=& 0,\\
 \label{eq:g2b}\Theta &=&
\frac{\dot{B}}{B}+2
\frac{\dot{Y}}{Y},\\
\label{eq:g2c}\sigma &=& \frac{1}{3
}\left(\frac{\dot{Y}}{Y}-\frac{\dot{B}}{B}\right),
\end{eqnarray}
\end{subequations}
respectively, and  dots denote the differentiation with respect to
$t$. The energy momentum tensor for the interior matter
distribution is described by
\begin{equation}
\label{eq:g3}T_{ab}= (\rho+p)u_a u_b +pg_{ab}+ {\pi}_{ab}+q_au_b
+q_bu_a,
\end{equation}
where $p$ is the isotropic pressure, $\rho$ is the energy density
of the fluid, ${\pi}_{ab}$ is the stress tensor and $q_a$ is the
heat flux vector. 
Anisotropy plays a significant role in gravitational collapse and affects the mass, luminosity 
and stability of relativistic spheres; these features have been highlighted in the
treatments of Chan \cite{C} and Herrera and Santos \cite{H}.
The stress tensor has the form
\begin{equation}
\label{eq:g4}{\pi}_{ab}= (p_r -p_t)\left(n_an_b - \frac{1}{3}
h_{ab}\right),
\end{equation}
where $p_r$ is the radial pressure, $p_t$ is the tangential
pressure and  $\mathbf{n}$ is a unit radial vector given by $n^a
=\frac{1}{B}\delta_1^a$. The isotropic pressure is given by
\begin{equation}
\label{eq:g5}p=\frac{1}{3} (p_r+2p_t)
\end{equation}
in terms of the radial pressure and the tangential pressure. For
the line element (\ref{eq:g1}) and matter distribution
(\ref{eq:g3}) the Einstein field equations become
\begin{subequations}
\begin{eqnarray}
\label{eq:g6a} \rho &=& 2
\frac{\dot{B}}{B}\frac{\dot{Y}}{Y}+\frac{1}{Y^2}+
\frac{\dot{Y}^2}{Y^2}- \frac{1}{B^2}\left(2 \frac{Y''}{Y}+
\frac{{Y'}^2}{Y^2}- 2
\frac{B'}{B}\frac{Y'}{Y}\right),\\
\label{eq:g6b} p_r &=& -2 \frac{\ddot{Y}}{Y}
-\frac{\dot{Y}^2}{Y^2}
- \frac{1}{Y^2}+\frac{1}{B^2} \frac{{Y'}^2}{Y^2},\\
\label{eq:g6c} p_t &=& - \left(\frac{\ddot{B}}{B}
+\frac{\dot{B}}{B}\frac{\dot{Y}}{Y}+\frac{\ddot{Y}}{Y}\right)
+\frac{1}{B^2} \left( \frac{Y''}{Y}-\frac{B'}{B}\frac{Y'}{Y}
\right),\\
\label{eq:g6d}q&=&
-\frac{2}{B^2}\left(\frac{\dot{B}}{B}\frac{Y'}{Y}-\frac{\dot{Y}'}{Y}\right),
\end{eqnarray}
\end{subequations}
where the heat flux $q^a =(0,q,0,0)$ is radially directed and primes
denote the differentiation with respect to $r$. The system of
equations (\ref{eq:g6a})-(\ref{eq:g6d}) governs the most general
situation in describing geodesic matter distributions in a
spherically symmetric gravitational field. These equations describe
the gravitational interaction of a shearing matter distribution with
heat flux and anisotropic pressure for particles traveling along
geodesics. From (\ref{eq:g6a})-(\ref{eq:g6d}), we observe that if
the gravitational potentials $B(t,r)$ and $Y(t,r)$ are specified
then the expressions for the matter variables $\rho, p_r, p_t$ and
$q$ follow by simple substitution.

The Vaidya exterior spacetime of radiating star is given by
\begin{equation}
\label{eq:g7} ds^2 = - \left(1-\frac{2m(v)}{R}\right)dv^2 -2dvdR+
R^2 (d{\theta^2}+\sin^2\theta d{\phi}^2),
\end{equation}
where $m(v)$ denotes the mass of the fluid as measured by an
observer at infinity. 
The line element (\ref{eq:g7}) is utilized to
describe incoherent null radiation which flows in the radial direction
relative to the hypersurface $\Sigma$ which denotes the boundary of
the star. The matching of the interior spacetime (\ref{eq:g1}) with
the exterior spacetime (\ref{eq:g7}) generate the set of junction
conditions on the hypersurface $\Sigma$ given by
\begin{subequations}
\begin{eqnarray}
\label{eq:g8a} dt &=&\left(1- \frac{2m}{R_{\Sigma}}+
2 \frac{dR_{\Sigma}}{dv}\right)^{\frac{1}{2}}dv,\\
\label{eq:g8b}Y(R_{\Sigma},t) &=& R_{\Sigma}(v),\\
\label{eq:g8c} m(v)_{\Sigma}&=& \left[\frac{Y}{2}\left(1+
\dot{Y}^2-\frac{{Y'}^2}{B^2}\right)\right]_{\Sigma},\\
\label{eq:g8d} (p_r)_{\Sigma} &=&(qB)_{\Sigma}.
\end{eqnarray}
\end{subequations}
The nonvanishing of the radial pressure at the boundary $\Sigma$ is
reflected in equation (\ref{eq:g8d}). Equation (\ref{eq:g8d}) is an
additional constraint which has to be satisfied together with the
system of equations (\ref{eq:g6a})-(\ref{eq:g6d}).

The junction condition (\ref{eq:g8d}) in the case of shear-free
spacetimes was first derived by Santos \cite{4}, and later it was
extended by Glass \cite{20} to incorporate spacetimes with nonzero
shear. On substituting (\ref{eq:g6b}) and (\ref{eq:g6d}) in
(\ref{eq:g8d}) we obtain
\begin{equation}
\label{eq:g9} 2 Y \ddot{Y}+\dot{Y}^{2}- \frac{{Y'}^2}{B^2}+
\frac{2}{B}Y\dot{Y}'  -2 \frac{\dot{B}}{B^2} Y Y' +1=0
\end{equation}
which has to satisfied on $\Sigma$. Equation (\ref{eq:g9}) governs
the gravitational behaviour of the radiating anisotropic star with
nonzero shear and no acceleration. As equation (\ref{eq:g9}) is
highly nonlinear, it is difficult to solve without some simplifying
assumption. This equation comprises two unknown functions $B(t,r)$
and $Y(t,r)$. To generate a solution we have to specify one of the
functions so that the resulting equation is tractable.

\section{New exact solutions}
For convenience we rewrite equation (\ref{eq:g9}) in the form of the
Riccati equation in the gravitational potential $B$ as follows
\begin{equation}
\label{eq:g10}
\dot{B}=\left[\frac{\ddot{Y}}{Y'}+\frac{\dot{Y}^2}{2YY'}
+\frac{1}{2YY'}\right]B^2 + \frac{\dot{Y}'}{Y'}B - \frac{Y'}{2Y}.
\end{equation}
Equation (\ref{eq:g10}) was analyzed by Nogueira and Chan \cite{21} who
obtained approximate solutions using numerical techniques. To
properly describe the physical features of a radiating relativistic
star exact solutions are necessary, preferably written in terms of
elementary functions. An exact solution was found by Naidu \emph{et
al} \cite{17} which was singular at the stellar centre. A new class of
solutions was established by Rajah and Maharaj \cite{18}, which contain
the Naidu \emph{et al} \cite{17} models, in which the singularities at
the centre were shown to be avoidable.

The Riccati equation (\ref{eq:g10}), which has to be satisfied on
the stellar boundary $\Sigma$, is highly nonlinear and difficult to
solve. Rajah and Maharaj \cite{18} obtained solutions in an {\it ad hoc}
fashion by assuming that the gravitational potential $Y(t,r)$ is a
separable function and specifying the temporal evolution of the
model. In this paper we demonstrate that it is possible to find new
exact solutions systematically without assuming separable forms for
$Y(t,r)$ and not fixing the temporal evolution of the model {\it a
priori}. If we introduce the transformation
\begin{equation}
\label{eq:g11} B= Z Y'
\end{equation}
then equation (\ref{eq:g10}) becomes
\begin{equation}
\label{eq:g12} \dot{Z}=\frac{1}{2Y}\left[F Z^2 -1\right],
\end{equation}
where we have set
\[F=2Y \ddot{Y}+\dot{Y}^2 +1.
\]
 We observe that equation (\ref{eq:g12}) becomes a separable
equation in $Z$ and $t$, and therefore integrable, if we let $F$ be
a constant or a function of $r$ only. In other word, (\ref{eq:g12})
is integrable as long as $F$ is independent of $t$. We emphasize
that in this approach we have not made any assumption about the
separability of the metric coefficients $B(t,r)$ and $Y(t,r)$ or
restricted the $t-$dependence. We demonstrate that this approach
leads to two new classes of solutions in the following sections.

\subsection{Analytic solution I}
If we set $F=1$ then the function $Y$ is given by
\begin{equation}
\label{eq:g13} Y(r,t)=[R_1(r) t +R_2(r)]^{2/3},
\end{equation}
where $R_1(r)$ and $R_2(r)$ are arbitrary functions of $r$. For
this case equation (\ref{eq:g12}) becomes
\begin{equation}
\label{eq:g14}  \dot{Z}=\frac{1}{2[R_1(r) t +R_2(r)]^{2/3}}\left[Z^2
-1\right].
\end{equation}
On integrating (\ref{eq:g14}) we obtain
\begin{equation}
\label{eq:g15}Z=\frac{1+ f(r) \exp
\left[3(R_1t+R_2)^{1/3}/R_1\right]} {1- f(r) \exp
\left[3(R_1t+R_2)^{1/3}/R_1\right]},
\end{equation}
 where $f(r)$ is a function of integration.
Hence from (\ref{eq:g11}), (\ref{eq:g13}) and (\ref{eq:g15}) we
get
\begin{equation}
\label{eq:g16}B=\frac{2}{3}\left[\frac{1+ f(r) \exp
\left[3(R_1t+R_2)^{1/3}/R_1\right]} {1- f(r) \exp
\left[3(R_1t+R_2)^{1/3}/R_1\right]}\right]\frac{[R_1't+R_2']}{[R_1
t+R_2]^{1/3}}.
\end{equation}
Therefore the line element (\ref{eq:g1}) takes the particular form
\begin{eqnarray}
\label{eq:g17} ds^2& =& - dt^2 +\frac{4}{9}\left[\frac{1+ f(r) \exp
\left[3(R_1t+R_2)^{1/3}/R_1\right]} {1- f(r) \exp
\left[3(R_1t+R_2)^{1/3}/R_1\right]}\right]^2\frac{[R_1't+R_2']^2}{[R_1
t+R_2]^{2/3}} dr^2 \nonumber\\
&&+[R_1(r) t +R_2(r)]^{4/3}(d\theta^2 +\sin^2 \theta d\phi^2),
\end{eqnarray}
The line element (\ref{eq:g17}) is given in terms of arbitrary
functions $R_1(r), R_2(r)$ and $f(r)$  so that it is possible to
generate infinite number of solutions for different choices of these
functions. Observe that if we set
\[R_1 =0, R_2 =r^{3/2}\]
then the equivalent of (\ref{eq:g17}) is
\[ds^2=-dt^2+dr^2+r^2(d{\theta}^2 + {\sin}^2 \theta d{\phi}^2)\]
which is the flat Minkowski spacetime. As the curvature then
vanishes we require $R_1 \neq 0$ and the $t$-dependence in $Y$
is maintained.

It is interesting to see that for particular forms  of the
arbitrary functions we regain  models  found previously. If we set
\[R_1=R^{3/2}, ~ R_2=a R^{3/2}\]
then the line element (\ref{eq:g17}) reduces to
\begin{equation}
\label{eq:g18} ds^2 = - dt^2 +(t+a)^{4/3}\left\{{R'}^2\left[\frac{1+
f(r) \exp \left[3(t+a)^{1/3}/R \right]} {1- f(r) \exp
\left[3(t+a)^{1/3}/R \right]}\right]^2dr^2+R^2(d\theta^2 +\sin^2
\theta d\phi^2)\right\}.
\end{equation}
The line element (\ref{eq:g18}) corresponds to the first category
of the Rajah and Maharaj \cite{18} models for an anisotropic radiating
star with shear. Furthermore note that if we set $a=0$ and $R=r$
then (\ref{eq:g18}) reduces to
\begin{equation}
\label{eq:ggg}ds^2 = - dt^2 +t^{4/3}\left\{\left[\frac{1+ f(r)
\exp \left[3t^{1/3}/r \right]} {1- f(r) \exp \left[3t^{1/3}/r
\right]}\right]^2dr^2+R^2(d\theta^2 +\sin^2 \theta
d\phi^2)\right\}.
\end{equation}
The metric (\ref{eq:ggg}) was first found by Naidu \emph{et al}
\cite{17} in their analysis of pressure anisotropy and heat
dissipation in a spherically symmetric radiating star undergoing
gravitational collapse. Note that when
\[R_1=r^{3/2},  R_2=0,~ f(r)=0\]
the line element (\ref{eq:g17}) takes on the simple form
\begin{equation}
\label{eq:g19} ds^2 = - dt^2 +t^{4/3}\left[dr^2+r^2(d\theta^2
+\sin^2 \theta d\phi^2)\right].
\end{equation}
The line element (\ref{eq:g19}) corresponds to the Friedmann
metric when the fluid is in the form of dust with vanishing heat
flux.

\subsection{Analytic solution II}
If we set $F=1+R_1^2(r)$ then the function $Y$ is given by
\begin{equation}
\label{eq:g20} Y(r,t)=R_1(r) t +R_2(r),
\end{equation}
where $R_1(r)$ and $R_2(r)$ are functions of $r$ only. For this
case equation (\ref{eq:g12}) becomes
\begin{equation}
\label{eq:g21} \dot{Z}=\frac{[R_1^2 +1]}{2[R_1 t+R_2]}\left[Z^2-
\frac{1}{[R_1^2 +1]}\right].
\end{equation}
The solution of (\ref{eq:g21}) can be written as
\begin{equation}
\label{eq:g22}Z= \frac{1}{\sqrt{R_1^2 +1}}\left[ \frac{1+g(r)[R_1
t+R_2]^{\sqrt{R_1^2 +1}/R_1}}{1-g(r)[R_1 t+R_2]^{\sqrt{R_1^2
+1}/R_1}}\right],
\end{equation}
where $g(r)$ is the function of integration. Hence from
(\ref{eq:g11}), (\ref{eq:g20}) and (\ref{eq:g22}) we get
\begin{equation}
\label{eq:g23}B=\frac{1}{\sqrt{R_1^2 +1}}\left[ \frac{1+g(r)[R_1
t+R_2]^{\sqrt{R_1^2 +1}/R_1}}{1-g(r)[R_1 t+R_2]^{\sqrt{R_1^2
+1}/R_1}}\right][R_1't+R_2'].
\end{equation}
Therefore the line element (\ref{eq:g1}) takes the particular form
\begin{eqnarray}
\label{eq:g24} ds^2& =& - dt^2 +\frac{1}{[R_1^2 +1]}\left[
\frac{1+g(r)[R_1 t+R_2]^{\sqrt{R_1^2 +1}/R_1}}{1-g(r)[R_1
t+R_2]^{\sqrt{R_1^2
+1}/R_1}}\right]^2[R_1't+R_2']^2dr^2 \nonumber\\
&&+[R_1(r) t +R_2(r)]^2(d\theta^2 +\sin^2 \theta d\phi^2)
\end{eqnarray}
in terms of arbitrary functions $R_1(r), R_2(r)$ and $g(r)$.
Therefore it is again possible to generate an infinite number of
solutions to (\ref{eq:g9}). As in \S3.1 we cannot set $R_1 =0$
became the Minkowski spacetime is results.

The new class of solutions given above does contain previously
known models. Note that when
\[R_1=R, ~ R_2=aR\]
the line element (\ref{eq:g24}) reduces to
\begin{equation}
\label{eq:g25} ds^2 = - dt^2 +(t+a)^2 \left\{\frac{{R'}^2}{[R^2
+1]}\left[ \frac{1+h(r)[t+a]^{\sqrt{R^2 +1}/R}}{1-h(r)[
t+a]^{\sqrt{R^2 +1}/R}}\right]^2 dr^2+R^2(d\theta^2 +\sin^2 \theta
d\phi^2)\right\},
\end{equation}
where we have defined the new arbitrary function $h(r)=
g(r)R^{\sqrt{R^2 +1}/R}$. The line element (\ref{eq:g25})
corresponds to the second category of the  Rajah and Maharaj \cite{18}
anisotropic radiating stars with shear. Such solutions are difficult
to interpret but could play a role in gravitational collapse for
strong fields.

\section{Discussion}
The simple form of the exact solutions that have been generated in
our treatment make it possible to study the physical features of
the model such as luminosity, rate of collapse, particle
production, neutrino flux and temperature profiles. In particular
explicit forms for the causal temperature can be found utilizing
the Maxwell-Catteneo heat transport equation
\begin{equation}
\label{eq:g26} \tau h_a^{~b}\dot{q_b}+q_a=-\kappa
\left(h_a^{~b}{\nabla}_b T + T \dot{u_a}\right).
\end{equation}
This has been done for special cases by Naidu \emph{et al} \cite{17} and
Rajah and Maharaj \cite{18}. The causal temperature is well behaved in
the stellar interior, in particular the causal temperature is
everywhere greater than the acausal temperature. Other choices of
the metric functions in our new classes of solutions (\ref{eq:g17})
and (\ref{eq:g24}) generate similar behaviour and highlight the role
of inhomogeneity in dissipative processes. Consequently our new
general class of exact radiating stars is physically reasonable.

We have generated a new class of stellar models with shear and
expansion in geodesic motion. All solutions found previously arise
as special cases in our treatment. Previous analysis were {\it ad
hoc} and effectively required that one of the metric functions
should be a separable function. The resulting Riccati equation could
then be solved. In this work, we did not assume separability of the metric
functions and did not initially fix the temporal evolution of the
model which is different from earlier treatments of this problem.
Our more general approach enabled us to solve the Riccati equation
systematically; we found that the Riccati equation could be
transformed to a separable equation. Two classes of exact solutions
were explicitly demonstrated to the junction condition. These were
shown to contain the Naidu \emph{et al} \cite{17} and Rajah and Maharaj
\cite{18} metrics. The fundamental reason that new solutions are possible
in that we have relaxed the condition of separability in the metric
function $Y(t,r)$.

\section*{Acknowledgements}
ST thanks the National Research Foundation and the University of
KwaZulu-Natal for financial support, and is grateful to Eastern
University, Sri Lanka for study leave. SDM acknowledges that this
work is based upon research supported by the South African
Research Chair Initiative of the Department of Science and
Technology and the National Research Foundation.

\end{document}